\documentclass[11pt, a4paper]{article}
\pdfoutput=1

\usepackage{amsmath}
\usepackage{amsfonts}
\usepackage{amssymb}
\usepackage{graphicx, rotating}
\usepackage{epstopdf}
\usepackage{epsfig}
\usepackage{latexsym}
\usepackage{graphicx}
\usepackage{color}
\usepackage{amsmath,bm,amssymb}
\usepackage{cite}
\usepackage{slashed}
\usepackage{hyperref}
\usepackage[utf8]{inputenc}
\hypersetup{colorlinks, citecolor=bluscuro, linkcolor=black, urlcolor=bluscuro}
\definecolor{rossos}{cmyk}{0,1,1,0.55}
\definecolor{bluscuro}{rgb}{0.15, 0.2, .85}
\definecolor{bluchiaro}{cmyk}{1,.3,0.,0.1}
\definecolor{pardo}{rgb}{0.15, 0.5, .15}

\newcommand{\negro}{\color{black}}

\newcommand{\be}{\begin{equation}}
\newcommand{\ee}{\end{equation}}
\newcommand{\bea}{\begin{eqnarray}}
\newcommand{\eea}{\end{eqnarray}}

\def\simgt{\stackrel{>}{{}_\sim}}

\newcommand{\GeV}{\,\mathrm{GeV}}

 \def\bea{\begin{eqnarray}}
  \def\eea{\end{eqnarray}}

\newcommand{\mH}{m_{H^0}}
\newcommand{\mA}{m_{A}}
\newcommand{\mHc}{m_{H^\pm}}
\newcommand{\tanb}{\tan\beta}

\newcommand{\mchi}{m_{\chi_1^0}}

\newcommand{\like}{{\cal L}}

\newcommand{\relic}{\Omega_\mathrm{DM} h^2}
\newcommand{\Omegaobs}{\Omega_{\rm DM}^\mathrm{obs}}
\newcommand{\Omegachi}{\Omega_{\chi_1^0}}

\begin{document}




\begin{titlepage}
\begin{flushright}
IFT-UAM/CSIC-19-160
\end{flushright}
\begin{center} ~~\\
\vspace{0.5cm} 
\Large {\bf\Large Generalized Blind Spots for Dark Matter Direct Detection in the 2HDM} 
\vspace*{1.5cm}

\normalsize{
{\bf  M. E. Cabrera$^a$\footnote[1]{meugenia@ecfm.usac.edu.gt},
J. A. Casas$^{b,c}$\footnote[2]{j.alberto.casas@gmail.com},
 A. Delgado$^c$\footnote[3]{antonio.delgado@nd.edu}
 and S. Robles$^d$\footnote[4]{sandra.robles@unimelb.edu.au}} \\

\smallskip  \medskip
$^a$\emph{Instituto de Investigaci\'on en Ciencias F\'isicas y Matem\'aticas, ICFM-USAC,\\ Universidad de San Carlos de Guatemala}\\
$^b$\emph{Instituto de F\'\i sica Te\'orica, IFT-UAM/CSIC,}\\
\emph{Universidad Aut\'onoma de Madrid, Cantoblanco, 28049 Madrid, Spain}\\
$^c${\it Department of Physics, University of Notre Dame,}\\
{\it Notre Dame, IN 46556, USA}}\\
$^d$\emph{ARC Centre of Excellence for Particle Physics at the Terascale \\
School of Physics, The University of Melbourne, Victoria 3010, Australia}

\medskip

\vskip0.6in 

\end{center}

\centerline{\large\bf Abstract}
\vspace{.5cm}
\noindent
In this paper we study the presence of generalized blind spots, i.e. regions of the parameter space where the spin-independent 
cross section for dark matter direct detection is suppressed, in the context of a generic 2HDM and a minimal fermionic Higgs-portal dark sector.  To this end, we derive analytical  expressions for the couplings of the dark matter to the light and heavy Higgses, and thus for the blind spot solutions. 
Unlike the case of a standard Higgs sector, blind spots can occur even without a cancellation between different contributions, 
 while keeping unsuppressed and efficient the annihilation processes in the early Universe involving Higgs states. 
As a consequence, the allowed parameter space is dramatically enhanced.

\vspace*{2mm}
\end{titlepage}


\section{Introduction}
\label{sec:intro}

Models featuring stable weakly interacting massive particles (WIMPs) are still among the most popular explanations for dark matter (DM)  \cite{Bertone:2004pz}. However, the most appealing  scenarios of this kind are under strong pressure, especially from direct detection (DD) experiments. More precisely, if the required annihilation of  WIMP particles in the early Universe occurs through interactions with either the $Z$ or Higgs bosons (the so-called $Z-$ and Higgs-portals), which somehow represent the minimal DM-SM interactions, a large portion of the parameter space of the simplest models is already excluded (see e.g. refs.~\cite{Arcadi:2014lta, Alves:2015pea,Escudero:2016gzx,Arcadi:2017kky, Gross:2017dan,Casas:2017jjg,Baum:2017enm,Arcadi:2019lka}). On the other hand, WIMP scenarios are still very attractive, not only for their  simplicity, but also because they naturally arise in well-motivated beyond-the-SM (BSM) physics models. 
Popular way-outs to this experimental pressure occur when the annihilation in the early Universe is enhanced, as it happens when it takes place in a resonant form (``funnels'') or through co-annihilation. These corners of the parameter space have attracted much attention, since they allow to save many interesting models of new physics.

There is another kind of regions of the parameter space where these economical WIMP scenarios can survive, namely when the DD elastic cross section is suppressed by some kind of cancellation. These are the so-called blind spots, which from the point of view of DM phenomenology offer a case of  similar interest to the above mentioned funnel and co-annihilation regions. Blind spots have been examined in the context of specific supersymmetric models \cite{Cheung:2012qy,Huang:2014xua,Anandakrishnan:2014fia,Crivellin:2015bva}, and in simplified (scalar or fermion) singlet-doublet DM models, i.e. when the dark sector contains a singlet and a doublet which can mix up \cite{Cheung:2013dua,
Berlin:2015wwa,
Banerjee:2016hsk, Arcadi:2018pfo}. 
The latter case is specially interesting, as it represents the relevant DM sector of many scenarios, such as generalized supersymmetric models and extra-dimensions.

 On the other hand, a usual feature of many BSM frameworks is the presence of an extended Higgs sector, typically containing two (or more) Higgs doublets. Actually, two-Higgs-doublet models (2HDM) have received much attention on their own for assorted physical reasons, see e.g.  ref.~\cite{Branco:2011iw,Jiang:2019soj}. Thus, it is natural to wonder how the appearance of blind spots changes when the Higgs sector is a 2HDM. The primary purpose of this paper is precisely to study this kind of scenario, deriving analytical expressions for the corresponding effective interactions of the WIMPs with ordinary matter, and thus for the blind-spot solutions. This will allow us to understand the general anatomy of the blind-spot regions for the different types of 2HDMs. 
As we will see, the presence of the second Higgs enhances the size of the blind spots not only because there are new ways to obtain a cancellation of the spin-independent DD cross section  (as already noticed in refs.~\cite{Huang:2014xua, Berlin:2015wwa,Arcadi:2018pfo}). Even without cancellations, the DD cross section can be small while keeping efficient the DM annihilation processes in the early Universe involving Higgs states. This leads to a dramatic enhancement of the viable regions of the parameter space.

The rest of the paper is summarized as follow. In section~\ref{sec:themodel} we introduce the model. In section~\ref{sec:anal} we provide analytical expressions for the different DM couplings, and thus for the generalized blind-spot condition. Section~\ref{sec:blind} is devoted to the appearance of the blind-spot regions in the alignment limit.
Finally, our conclusions are presented in section~\ref{sec:conclu}.

\section{The model}
\label{sec:themodel}

As mentioned in the introduction, we will focus on a model with two Higgs doublets, $\Phi_1$, $\Phi_2$, with hypercharges $Y=1/2$ (our notation here follows that of ref.~\cite{Bernon:2015qea}). The dark sector consists of two fermionic $SU(2)$ doublets $D_1, D_2$ ($\sim$ ``Higgsinos''),  with hypercharges $+1/2$, $-1/2$ respectively, and a fermionic singlet  $S$ ($\sim$ ``Singlino'') with zero hypercharge. The $D_1, D_2$ states can be combined in a Dirac fermion, if desired. The notation here follows that of ref.~\cite{Banerjee:2016hsk}, in order to facilitate comparisons.  Note that this Dirac fermion represents the minimal UV completion of a fermion-singlet Higgs-portal scenario for DM. 

The relevant terms of the most general Lagrangian for the dark sector are:
\be
-{\cal L}\supset \frac{1}{2}M_S SS + M_D D_1D_2 + y_1^1 SD_1\bar \Phi_1 + y_2^1 SD_2 \Phi_1 + y_1^2 SD_1\bar \Phi_2
+ y_2^2 SD_2\Phi_2 \ +\ h.c. \ ,
\label{lag}
\ee
where $\bar \Phi_{1,2}= i\sigma_2 \Phi_{1,2}^*$ and $M_D, M_S$ ($y_i^j$) are mass (Yukawa-coupling) parameters. The Higgs doublets acquire vacuum expectation values (VEVs) according to the structure of the 2HDM Higgs potential, which we do not write here explicitly (for further details see ref.~\cite{Bernon:2015qea}). Then, the CP-even neutral part of the Higgses reads
\be
\Phi_{1,2} = \frac{1}{\sqrt{2}}
\left(\begin{array}{c} 0\\v_{1,2}+h_{1,2}^0  \end{array}   \right),\ \ 
\bar \Phi_{1,2} = \frac{1}{\sqrt{2}}
\left(\begin{array}{c} v_{1,2}+h_{1,2}^0\\0  \end{array}   \right),\ \ 
\label{h120}
\ee
where $v_1^2+v_2^2=v^2= (246 \ {\rm GeV})^2$. As usual, we define the $\tan \beta$ parameter so that
\be
v_1=v\ \cos\beta,\ \ v_2= v\  \sin\beta .
\ee
With these definitions, the relevant Lagrangian for the neutral states reads
\bea
-{\cal L}&\supset& 
\frac{1}{2}M_S SS - M_D D_1^0D_2^0 - \frac{1}{\sqrt{2}}y_1^1 SD_1^0(v_1+h_1^0) + \frac{1}{\sqrt{2}}y_2^1 SD_2^0(v_1+h_1^0) 
\nonumber \\
 && -\frac{1}{\sqrt{2}}\ y_1^2 SD_1^0(v_2+h_2^0)
+ \frac{1}{\sqrt{2}}y_2^2 SD_2^0(v_2+h_2^0) \ +\ h.c.
\label{lag2}
\eea
Although in principle all the masses and couplings in the Lagrangian are complex, there are only three independent phases, namely those of $M_S^*M_D^*y_1^1y_2^1$, $y_1^1(y_1^2)^*$, $y_2^1(y_2^2)^*$, which we will assume to be real to avoid CP violations. This allows to take the six parameters of the Lagrangian as real quantities and to fix the sign of three of them. We will make use of this freedom later.

Of course, the fields appearing in the previous expression do not correspond to the mass eigenstates. For the Higgs sector the latter are  $h^0, H^0$, i.e. the light (standard)
and the heavy Higgses, respectively, which are related to the original $h_1^0, h_2^0$ fields by a basis rotation
\be
\left(\begin{array}{c} h^0\\H^0 \end{array}   \right) = 
 \left(\begin{array}{cc} \cos\alpha & -\sin\alpha \\ 
\sin\alpha & \cos\alpha  \end{array}\right)
\left(\begin{array}{c} h_2^0\\h_1^0  \end{array}   \right).
\label{alpha_rot2} 
\ee
The masses of the two physical Higgses ($m_{h^0}=125$ GeV, $\mH$) are determined by the structure of the Higgs potential. This is also true for the $\alpha$ angle, which, in principle, is independent of $\beta$; although in the decoupling limit ($\mH\rightarrow\infty$) they are related by $\alpha = \beta-\pi/2$.

Concerning the (neutral) fermionic sector, the ``neutralino'' mass eigenstates, $\chi_{1,2,3}^0$ arise upon diagonalization of the mass matrix ${\cal M}_N$, defined as
\be
-{\cal L}_{\rm mass}= \frac{1}{2} (S, D_1^0, D_2^0)\ {\cal M}_N\ 
\left(\begin{array}{c} S\\D_1^0\\D_2^0  \end{array}   \right)\ +\ h.c.\ ,
\ee
where, from Eq.~(\ref{lag2}),
\be
{\cal M}_N\ =
\left(
\begin{array}{ccc} M_S & -\frac{1}{\sqrt{2}}\left(y_1^1 v_1 + y_1^2 v_2 \right) & 
\frac{1}{\sqrt{2}} \left(y_2^1 v_1 + y_2^2 v_2 \right) \\
-\frac{1}{\sqrt{2}} \left(y_1^1 v_1 + y_1^2 v_2 \right) & 0 & -M_D \\
\frac{1}{\sqrt{2}} \left(y_2^1 v_1 + y_2^2 v_2 \right) & -M_D & 0 
\end{array}
\right)\ .
\label{MN}
\ee

Taking appropriate limits on the various parameters, we can recover simpler scenarios. In particular the relevant dark sector is a pure ``Singlino'' (``Higgsino'') for $M_D\rightarrow \infty$ ($M_S\rightarrow \infty$). Similarly, the scenario of just one (SM) Higgs is recovered for $\mH\rightarrow\infty$, $\alpha = \pi/2 -\beta$, as mentioned above.

\section{Analytical expressions for DM couplings and blind spots}
\label{sec:anal}
We are interested in the effective couplings, $y_{h \chi_i\chi_i}$, $y_{H \chi_i\chi_i}$ of the fermion mass  eigenstates, $\chi_{i}^0$ (in particular $\chi_{1}^0$, which is the DM particle) to the physical Higgs states, $h^0, H^0$, i.e.
\be
-{\cal L}\supset y_{h \chi_i\chi_i} h^0\chi_i^0 \chi_i^0 + y_{H \chi_i\chi_i} H^0\chi_i^0 \chi_i^0\ .
\label{ydef}
\ee
Using Eq.~(\ref{alpha_rot2}), these can be written in terms of the analogous couplings in the initial basis, namely
\bea
y_{h \chi_i\chi_i}&=& -s_\alpha\  y_{h_1 \chi_i\chi_i} \ +\ c_\alpha\ y_{h_2 \chi_i\chi_i},
\label{hcoupling}\\
y_{H\chi_i\chi_i}&=& \ \ c_\alpha\ y_{h_1 \chi_i\chi_i} \ +\ s_\alpha\  y_{h_2 \chi_i\chi_i},
\label{Hcoupling}
\eea
where we have used the shorthand $s_\phi=\sin \phi$, $c_\phi=\cos\phi$.
Now, from Eqs.~(\ref{lag2}, \ref{MN}), the $y_{h_{1,2} \chi_i\chi_i}$ couplings can be written as 
\be
y_{h_{a}\chi_i\chi_i}=\pm \frac{1}{2} \frac{\partial m_{\chi_i^0}}{\partial v_a}
,\ \ \ \ \ a=1,2 \ ,
\label{yhi}
\ee
where the $\pm$ sign corresponds to the case where the $m_{\chi_i}$ eigenvalue of the mass matrix is positive or negative. On the other hand, an analytical expression for ${\partial m_{\chi_i^0}}/{\partial v_a}$, and thus for $y_{h_{a}\chi_i\chi_i}$, can be obtained from the eigenvalue equation,
\bea
\frac{\partial}{\partial v_a} \left|  {\cal M}_N - m_{\chi_i^0} I  \right| = 0.
\eea
More precisely,
\be
y_{h_{a}\chi_i\chi_i}= -\frac{1}{2}\ \frac{{\cal N}_{a}}{{\cal D}};\ \ \ \ \ a=1,2 \ ,
\label{yhi2}
\ee
with
\bea
{\cal N}_a &=&  \pm M_D \left[y_1^a \left(y_2^1 v_1 + y_2^2 v_2 \right) + y_2^a
\left(y_1^1 v_1 + y_1^2 v_2 \right)
\right]
\nonumber\\
&+& |m_{\chi_i^0}|  \left[y_1^a \left(y_1^1 v_1 + y_1^2 v_2 \right)+ y_2^a
\left(y_2^1 v_1 + y_2^2 v_2 \right) 
\right],
\eea
and 
\be
{\cal D} = \pm2|m_{\chi_i^0}|M_S -3m_{\chi_i^0}^2+\frac{1}{2} \left[ \left(y_1^1 v_1 + y_1^2 v_2 \right)^2+\left(y_2^1 v_1 + y_2^2 v_2 \right)^2
\right] + M_D^2 \ .
\label{D}
\ee
Plugging Eq.~(\ref{yhi2}) into Eqs.~(\ref{hcoupling}, \ref{Hcoupling}), we obtain analytical expressions for the couplings of $h^0$ and $H^0$ to the mass eigenstates:
\bea
y_{h \chi_i\chi_i}=-\frac{1}{2{\cal D}}
\left\{  \pm M_D \left[y_1 \left(y_2^1 v_1 + y_2^2 v_2 \right) + y_2
\left(y_1^1 v_1 + y_1^2 v_2 \right)
\right] \right.
\nonumber\\
\left. + |m_{\chi_i^0}|  \left[y_1 \left(y_1^1 v_1 + y_1^2 v_2 \right)+ y_2
\left(y_2^1 v_1 + y_2^2 v_2 \right) 
\right] \right\},
\label{yhXX}
\eea
\bea
y_{H \chi_i\chi_i}=-\frac{1}{2{\cal D}}
\left\{  \pm M_D \left[\tilde y_1 \left(y_2^1 v_1 + y_2^2 v_2 \right) + \tilde y_2
\left(y_1^1 v_1 + y_1^2 v_2 \right)
\right] \right.
\nonumber\\
\left. + |m_{\chi_i^0}|  \left[\tilde y_1 \left(y_1^1 v_1 + y_1^2 v_2 \right)+ \tilde y_2
\left(y_2^1 v_1 + y_2^2 v_2 \right) 
\right] \right\},
\label{yHXX}
\eea
where we have defined
\bea
y_1=&-y_1^1s_\alpha+y_1^2c_\alpha,\ \ &y_2= -y_2^1s_\alpha+y_2^2c_\alpha,
\nonumber \\
{\tilde y}_1=&y_1^1c_\alpha+y_1^2s_\alpha,\ \  &{\tilde y}_2 = y_2^1c_\alpha+y_2^2s_\alpha .
\label{ys}
\eea
Note that $y_{H^0 \chi_i\chi_i}$ can be obtained from $y_{h^0 \chi_i\chi_i}$ by simply replacing  $y_{1,2}\rightarrow \tilde y_{1,2}$. The reason is simply that $y_{1,2}$, $\tilde y_{1,2}$ are the couplings of $h^0$, $H^0$ to the initial $D_{1,2}$ doublets. Namely, from Eqs.~(\ref{lag2}, \ref{alpha_rot2}),
\bea
-{\cal L}\supset
-\frac{1}{\sqrt{2}}y_1 SD_1^0 h^0 + \frac{1}{\sqrt{2}}y_2 SD_2^0 h^0 - \frac{1}{\sqrt{2}}\tilde y_1 SD_1^0 H^0 + \frac{1}{\sqrt{2}}\tilde y_2 SD_2^0 H^0 \ +\ h.c. \ ,
\label{Lyytilde}
\eea
with $y_{1,2}$, $\tilde y_{1,2}$ given by Eq.~(\ref{ys}).

From expressions (\ref{yhXX}-\ref{ys}), it is straightforward to obtain the blind spots, i.e. the region of parameters where the spin-independent DD cross section is suppressed. Generically, the amplitude for the DM-nucleon scattering, $\chi_1^0 N \rightarrow \chi_1^0 N$, mediated by a Higgs ($h^0$ or $H^0$ in $t-$channel) is  proportional to the effective coupling, $y^{\rm eff}_{\rm DD}/m_h^2$, with
\bea
{\negro y^{\rm eff}_{\rm DD}}\equiv \sum_q \left[y_{h \chi_1\chi_1} +  \frac{m_h^2}{m_H^2} C_q\ y_{H \chi_1\chi_1}\right] f_q^N.
\label{BS}
\eea
Here $q$ runs over the quarks in the nucleon, $N$; 
$f_q^N$ (with $N=p,n$) are the hadronic matrix elements, determined either experimentally 
or by lattice QCD simulations and related to the mass fraction of $q$ within the nucleon; and $C_q$ is a numerical factor that gives the departure of the coupling of the quark $q$ to the heavy Higgs, $H^0$, from that of the SM Higgs, $h^0$. The $C_q$ factors depend on the type of the 2HDM considered (more details in the next section). 
Whenever ${\negro y^{\rm eff}_{\rm DD}}\simeq 0$,  we are in a blind spot region of the parameter space.

\section{Blind spots in the alignment limit}
\label{sec:blind}

Experimental constraints indicate that the $125$ GeV Higgs of a 2HDM cannot be very different from the SM Higgs \cite{Aad:2012tfa,Chatrchyan:2012ufa,Aad:2013wqa,Chatrchyan:2013lba,Sirunyan:2017exp,Aaboud:2018wps}.  We adopt the conservative approach that the light Higgs is $100\%$ SM-like (a situation which is usually called ``alignment''), which  effectively means that the heavy Higgs does not obtain a VEV, i.e. it is inert. The analysis of this model when the heavy Higgs is allowed to have a VEV consistent with the present experimental data will be postponed for a forthcoming paper.
Consequently, from now on we will concentrate on the alignment limit to illustrate the structure of the blind spots.

The exact alignment limit (with the light Higgs, $h^0$, playing the role of the SM Higgs boson) occurs for $c_{\beta-\alpha}=0$, i.e. $\alpha=\beta-\pi/2$. Then, we can recast expressions (\ref{yhXX}, \ref{yHXX}) as
\bea
y_{h \chi_i\chi_i}&=&-\frac{1}{2{\cal D}}y^2 v
\ (  \pm M_D \sin 2\theta + |m_{\chi_i^0}|) \ ,
\label{yhXXal} \\
y_{H \chi_i\chi_i}&=&-\frac{1}{2{\cal D}}y \tilde y v
\ (  \pm M_D \sin (\theta+\tilde\theta) + |m_{\chi_i^0}|\cos(\theta-\tilde \theta)) \ ,
\label{yHXXal}
\eea
with
\be
y^2 =(y_1)^2+(y_2)^2,\ \ \ c_\theta = \frac{y_1}{y}, \ s_\theta = \frac{y_2}{y}\ ,
\nonumber
\ee
\be
\tilde y^2 =(\tilde y_1)^2+(\tilde y_2)^2,\ \ \ c_{\tilde \theta} 
= \frac{\tilde y_1}{\tilde y}, \ s_{\tilde \theta} = \frac{\tilde y_2}{\tilde y}\ .
\ee
The expression for ${\cal D}$, Eq.~(\ref{D}), is also simplified
\be
{\cal D} = \pm2|m_{\chi_i^0}| M_S -3m_{\chi_i^0}^2+\frac{1}{2} y^2 v^2 + M_D^2 \ .
\label{D2}
\ee

 Note that $y_{h \chi_i\chi_i}\propto y^2$, while $y_{H \chi_i\chi_i}\propto y\tilde y$. This occurs because in the alignment limit the Higgs VEV comes entirely from the doublet associated with the light Higgs, $h^0$, which thus appears always in the $v+h^0$ combination. Then, due to $SU(2)$ invariance both effective couplings involve a $v-$factor and thereby a $y-$Yukawa, see Eq.~(\ref{Lyytilde}). For the same reason, it is easy to show that
 in this limit the expression (\ref{yhXXal}) for
 $y_{h \chi_i\chi_i}$ can be obtained from 
\be
y_{h \chi_i\chi_i}=\pm \frac{1}{2} \frac{\partial m_{\chi_i^0}}{\partial v}
.
\label{yh_al}
\ee

\vspace{0.2cm}
\noindent
In the following, we make use of the freedom to fix the sign of three parameters to take $M_S$, $y_1$ and $y_2$ as positive, while the signs of $M_D$, $\tilde y_1, \tilde y_2$ can be positive or negative. Consequently, $\theta\in[0,\pi/2]$, $\tilde \theta\in[0,2\pi]$. This convention allows to scan the whole parameter space in a complete and non-redundant way; and it converges to the sign convention used in ref.~\cite{Banerjee:2016hsk} for the case of a single Higgs.

\subsection{Alignment from decoupling}

A somewhat trivial way to obtain alignment is through decoupling, i.e. when $\mH\gg m_{h^0}$ (for details see ref.~\cite{Bernon:2015qea}). Then, the contribution of the heavy Higgs to the DD cross section becomes negligible and the effective coupling $y^{\rm eff}_{\rm DD}$ of Eq.~(\ref{BS}) 
reads
\bea
y^{\rm eff}_{\rm DD}\ \propto\ y_{h \chi_1\chi_1}\ \propto \ 
 \pm M_D \sin 2\theta + |\mchi|,
\label{BS2}
\eea
which agrees with the expression obtained in ref.~\cite{Banerjee:2016hsk} for just one Higgs, as expected. Note that in this limit a blind spot is only possible when $M_D<0$. The reason is the following. From Eq.~(\ref{yh_al}), the blind spot condition, $y_{h \chi_1\chi_1}=0$, implies that $\mchi$ does not depend on $v$, and thus must be equal to one of the mass eigenvalues of the mass matrix,  Eq.~(\ref{MN}), when $v=0$, i.e. $M_S$, $M_D$ or $-M_D$. However, since $\pm M_D \sin 2\theta + |\mchi|=0$, this can only be achieved (barring the  $\sin 2 \theta = 1$ case) if $\mchi=M_S$ (and thus positive) and $M_D<0$. Note also that in the decoupling limit the existence of a blind spot requires $M_S\leq|M_D|$, barring the aforementioned case.

\begin{figure}[ht]
    \centering
    \includegraphics[scale=0.55]{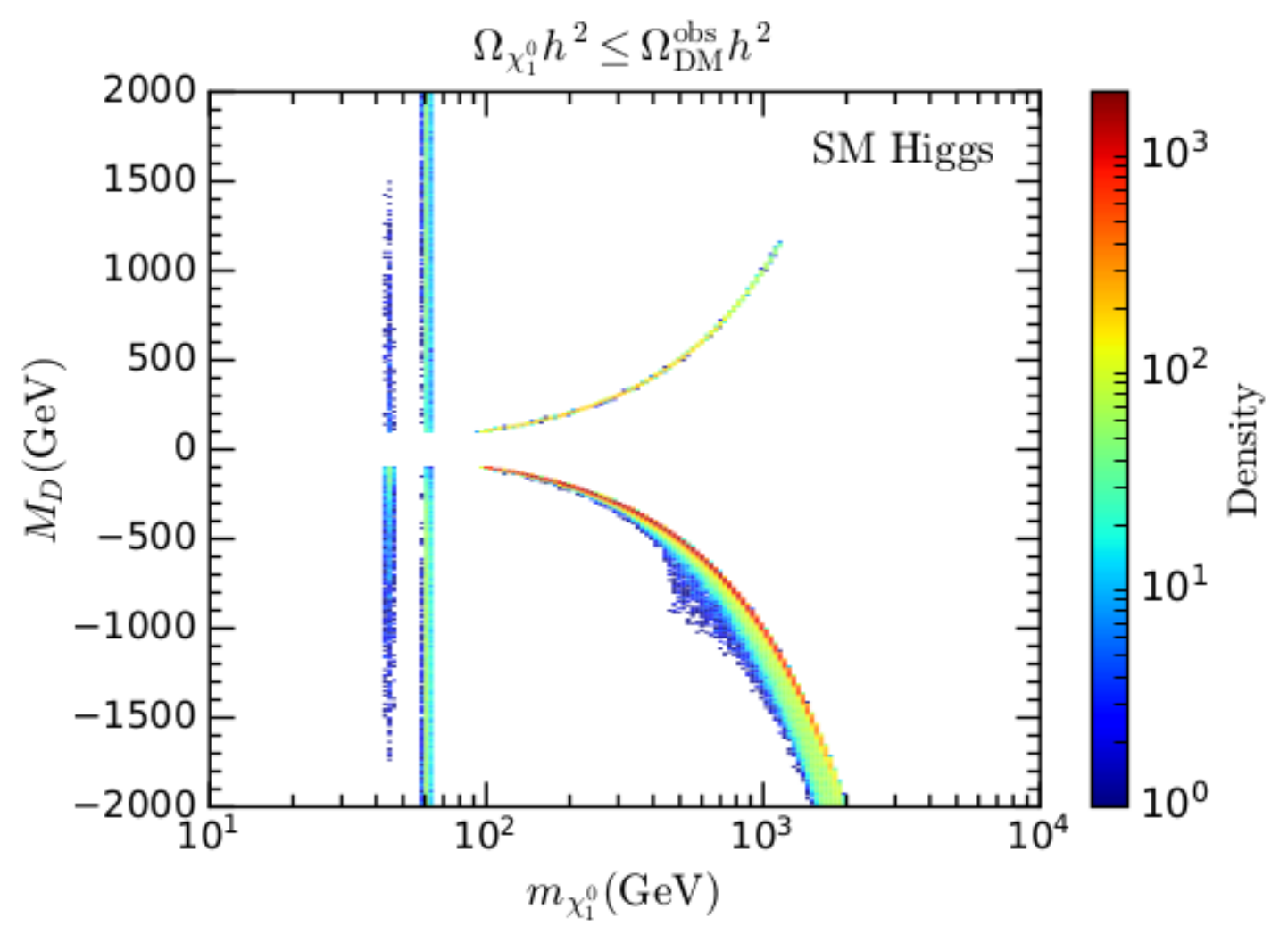}
    \caption{Regions of the parameter space allowing $\Omegachi h^2\leq \Omegaobs h^2$ in the decoupling limit (equivalent to a SM Higgs sector).  All the plotted points fulfill the bounds from XENON1T~\cite{Aprile:2018dbl,Aprile:2019dbj} and PICO-60~\cite{Amole:2019fdf}. Moreover the condition $|M_D|>100$ GeV has been required to satisfy the LEP limits \cite{Heister:2003zk} on charged fermions, in this case the charged components of the $D-$fields.
    The color code indicates the relative density of points. }
    \label{fig:type1H1}
\end{figure}

All this is illustrated in the scan of Fig.~\ref{fig:type1H1}  which shows the physically viable region in the $\mchi-m_D$ plane where $\Omegachi\leq \Omegaobs$, fulfilling DD bounds from XENON1T~\cite{Aprile:2018dbl,Aprile:2019dbj} and PICO-60~\cite{Amole:2019fdf}. 
The two narrow and dense strips at $\mchi=\pm M_D$ correspond to models where $\chi_1^0$ is either almost a pure doublet, i.e. a combination of the $D_1^0, D_2^0$ fields; or a well-tempered mixture of $S$ and $D_1^0, D_2^0$~\cite{ArkaniHamed:2006mb}. Comparison of the upper branch (where there is no blind spot and the dark matter is in a well-tempered regime) with the lower one shows the noticeable effect of the blind spot. 
The $Z-$ and $h-$funnel regions are also visible in the plot.

The results for $\Omegachi = \Omegaobs$ are very similar, except for the upper strip.  The reason is that when $\chi_1^0$ is a pure doublet, the annihilation is ``too efficient" except around 1 TeV, which is the only piece
of the upper strips that survives. On the other hand, in the well-tempered regime it is possible to obtain the correct
relic density with $M_S\sim |M_D| \sim \mchi$, but at the price of raising the
$y-$couplings in a way that DD excludes the model when $M_D>0$. 
In contrast, when $M_D<0$ the $y-$couplings can be arranged according to the blind spot relation, Eq.~(\ref{BS2}) $\simeq 0$, thus evading DD bounds. Hence the lower strip survives in this regime, albeit not as a particularly dense region.  Consequently, apart from the funnels and a narrow region at $|M_D|\simeq 1$ TeV, all the regions rescued for $\Omegachi =\Omegaobs$ correspond to the blind spot condition.

To perform the previous scan and those of the next subsection, we have implemented the model in {\tt FeynRules}~\cite{Christensen:2008py,Alloul:2013bka}, interfaced with {\tt CalcHEP}~\cite{Christensen:2009jx}. More specifically, we have extended the publicly available 2HDM model files~\cite{Degrande:2014vpa}, considering only tree level interactions, with a singlet fermion and two doublet fermions as described in section~\ref{sec:themodel}. Then, the relic abundance and the elastic scattering cross sections have been calculated with  {\tt microOMEGAS} \cite{Belanger:2006is}.  
The scan has been performed in the following ranges of the relevant parameters:
\be
M_S\in [10,2000]\ {\rm GeV},\ \ \ M_D\in [-2000,2000]\ {\rm GeV},\ \ \ 
y_1, y_2 \in [0,1], 
\ee
with a log prior on $M_S$ and flat priors on the remaining parameters, using {\tt MultiNest} for an efficient exploration of the parameter space~\cite{Feroz:2007kg,Feroz:2008xx,Feroz:2013hea}. 
To that end, we have constructed a joint likelihood function, as follows 
\begin{equation}
 \log \like_{\rm Joint} ~=~ \log\like_{\relic} + \log\like_{\rm Xenon1T}~, 
 \label{eq:jointlike}
\end{equation}
where $\like_{\relic}$ is implemented as an upper bound with a  smeared step-function \cite{deAustri:2006jwj}, centered at the observed value \cite{Aghanim:2018eyx}. $\like_{\rm Xenon1T}$ is calculated using {\tt RAPIDD} \cite{Cerdeno:2018bty}, a surrogate model for fast computation of the expected DM spectrum in direct detection experiments,  tuned to the latest XENON1T results \cite{Aprile:2018dbl,Aprile:2019dbj}. 
Here and throughout the paper the DD cross section has been weighted by the $\xi={\rm min}[1,\Omegachi/\Omegaobs]$ factor, which appropriately scales the cross section for under-abundant dark matter.

It is apparent from Fig.~\ref{fig:type1H1} that, apart from the funnels, the absolute value of $M_D$ cannot be very different from $M_S$, even in the blind spot region. The reason is the following. Since in the decoupling limit the blind spot condition is equivalent to $y_{h \chi_1\chi_1}=0$, the coupling of the DM particle, $\chi_1^0$ to the SM Higgs vanishes for both cross sections DM-nucleon elastic scattering  and  DM annihilation in the early Universe. {\negro In particular, the processes $\chi_1^0\chi_1^0 \rightarrow h^0 \rightarrow {\rm SM\ SM}$, $\chi_1^0\chi_1^0 \rightarrow h^0\ h^0$ vanish at first order.} Consequently, in the blind spot the required annihilation occurs thanks to the $D-$component of $\chi_1^0$ and the corresponding weak interaction, {\negro as in the well-tempered regime.} Since this component is inversely proportional to the difference of masses, if $|M_D|$ is much larger than $M_S$, the effective weak coupling of $\chi_1^0$ becomes too small to provide  the required amount of DM annihilation. {\negro The results of this subsection are consistent with those obtained in ref. \cite{Banerjee:2016hsk}.}

\subsection{Alignment without decoupling. Blind spots without cancellations}

A more interesting case arises when the alignment is achieved without decoupling \cite{Carena:2013ooa,Bernon:2015qea}. This occurs whenever the coupling in the Higgs-potential denoted by $Z_6$ in ref.~\cite{Bernon:2015qea} is vanishing or very small. Then, still $\alpha=\beta-\pi/2$, and $\mH$ can be quite low without conflicting with any experimental constraints. The precise lower bound depends on the type of 2HDM under consideration \cite{Bernon:2015qea}.

In the Type I 2HDM, defined by the fact that one of the initial doublets, say $\Phi_2$, is the only one that couples to all fermions, the bounds are very mild. Actually, $\mH$ can be close to $m_{h^0}=125$ GeV without conflicting with experiments. In the Type II 2HDM, in which $\Phi_1$ couples to down-like quarks and charged leptons, and $\Phi_2$ to up-like quarks, just as in supersymmetry, the bounds are more restrictive. This is mainly due to the limits from $H^0\to \tau^+\tau^-$, since in the Type II the coupling of $H^0$ to charged leptons is enhanced by $\tan\beta$ (see below). Generically, taking $\mH\geq 400$ GeV is safe, although it can be much lower (even below 200 GeV) if $m_A\geq 400$ GeV \cite{Bernon:2015qea}. 

For DD matters, the most important difference between the Type I and  Type II 2HDMs concerns the couplings to quarks, which are given in Table~1 \cite{Bernon:2015qea}

\begin{table}[h]
    \centering
    \begin{tabular}{|c||c|c||c|c|}
    \hline
    \multicolumn{1}{|c||}{} &
    \multicolumn{2}{|c||}{Type I} & \multicolumn{2}{|c|}{Type II}\\
    \hline
      Higgs &
      $u-$quarks & $d-$quarks and leptons&  $u-$quarks & $d-$quarks and leptons \\
      \hline
      $h^0$ &	$\cos\alpha/\sin\beta$ & $\cos\alpha/\sin\beta$ & $\cos\alpha/\sin\beta$ & $-\sin\alpha/\cos\beta$\\
      $H^0$ &	$\sin\alpha/\sin\beta$ & $\sin\alpha/\sin\beta$ & $\sin\alpha/\sin\beta$ & $\cos\alpha/\cos\beta$\\
      \hline     
    \end{tabular}
    \caption{Factors for couplings of Higgs states to SM fermions in the Type I and Type II 2HDMs relative to those of the SM.}
    \label{tab:couplings}
\end{table}{}

In the alignment limit, $\alpha=\beta-\pi/2$ and we recover the SM couplings for the conventional Higgs, $h^0$. However, the couplings of the heavy Higgs, $H^0$, to $u-$ and $d-$quarks acquire the following factors
\bea
&C_u=-\cot \beta,&\ \ C_d=-\cot\beta\ \ \ \ \text{(Type I)},
\nonumber\\
&C_u=-\cot \beta,&\ \ C_d=\tan\beta\ \ \ \ \ \  \text{(Type II)}.
\eea
These are the $C_q$ coefficients to plug in  expression (\ref{BS}) for the blind spot condition. An important point is that, with two Higgs states in play, $ y^{\rm eff}_{\rm DD}$ can vanish not because the couplings of the DM particle to the Higgses vanish (unique possibility when there was just one Higgs), but because their contributions cancel in Eq.~(\ref{BS}). Consequently the blind spot condition can be accomplished and, simultaneously,  DM particles can efficiently annihilate in the early Universe thanks to sizeable interactions with both  Higgses. This opens enormously the available parameter space; in particular it is not necessary anymore that $|M_D|$ is close to $M_S$, as happened for the case of a unique Higgs
(barring funnels and the ``pure Higgsino" region). 

Actually, $y^{\rm eff}_{\rm DD}$ can be small {\em not} due to a cancellation between the various terms in Eq.~(\ref{BS}), but simply because all of them are small (i.e. with no need of tuning), and still the annihilation of DM involving Higgses can be efficient enough. To see this, note first that the $H^0$ contribution to the DD cross section, given by the second term of Eq.~(\ref{BS}), can be small not because the $\tilde y-$couplings are small but because the prefactor $\frac{m_h^2}{m_H^2} C_q$ is. This is the typical case for the Type I 2HDM, since $C_q=-\cot\beta$ and $\tan\beta \simgt 1$ to avoid a non-perturbative top Yukawa coupling. 
Then, if the $\tilde y-$couplings are sizeable, the processes $\chi_1^0\chi_1^0 \rightarrow H^0
\rightarrow {\rm SM\ SM}$, $\chi_1^0\chi_1^0 \rightarrow H^0\ H^0$,  $\chi_1^0\chi_1^0 \rightarrow Z\ H^0$ and others can be efficient enough to provide the required DM annihilation. More generically, even if the above prefactor is ${\cal O} (1)$, the $H^0$ contribution in Eq.~(\ref{BS}) can be small because the $y-$couplings (not the $\tilde y-$couplings) are small, as can be seen from the expression for  $y_{H\chi_1\chi_1}$,  Eq.~(\ref{yHXXal}). In that case,  processes like $\chi_1^0\chi_1^0 \rightarrow H^0\ H^0$ (proportional to $\tilde y^2$) can be equally efficient. 

The bottom line of the previous paragraph is that, due to the presence of the second Higgs,  the couplings involved in DM annihilation are not necessarily those involved in DD. This only happens for certain annihilation processes, as $\chi_1^0\chi_1^0 \rightarrow h^0\ h^0$. 
As a result, large ``blind spot'' regions that were unviable in the decoupling (or one-Higgs) limit are now rescued. Actually, they are blind spots only in the sense that $y^{\rm eff}_{\rm DD}$ is very small, but this does not necessarily imply a cancellation between contributions.

\begin{table}[ht]
    \centering
    \begin{tabular}{|c|c|c||c|c|c|}
    \hline
    \multicolumn{3}{|c||}{Type I} & \multicolumn{3}{|c|}{Type II}\\
    \hline
      $\mH(\GeV)$ & $\tanb$ & $\mA=\mHc(\GeV)$ &  $\mH(\GeV)$ & $\tanb$ & $\mA=\mHc(\GeV)$\\
      \hline
      300 & 5 & 600 & 300 & 5 & 600 \\
      300 & 30 & 600 & 300 & 30 & 600\\
800  & 5 & 800 & 800 & 5 & 800\\
      \hline     
    \end{tabular}
    \caption{Benchmarks for the Type I and Type II 2HDMs.}
    \label{tab:benchmarks}
\end{table}{}

\begin{figure}[ht]
    \centering
    \includegraphics[scale=0.55]{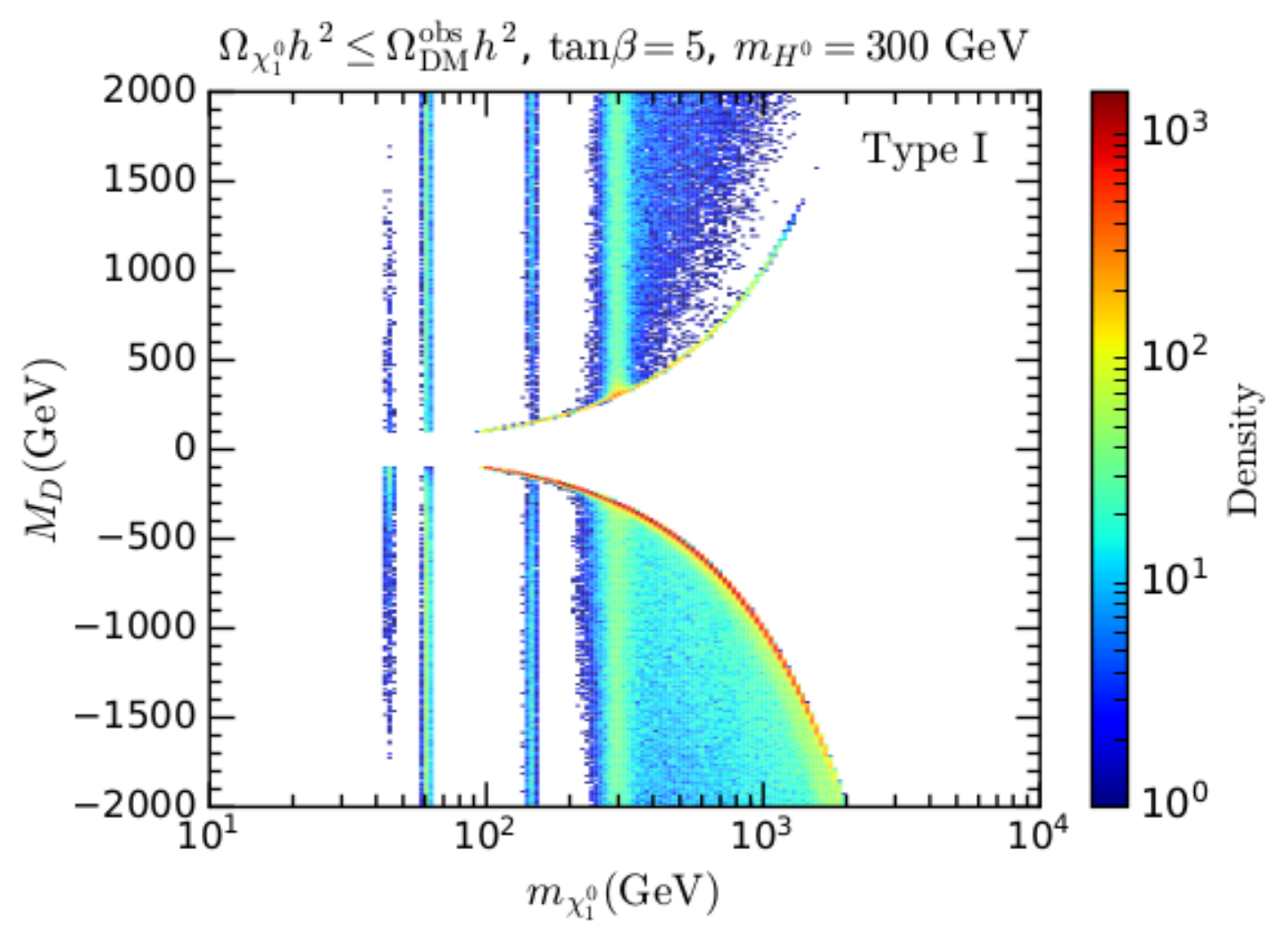}
    \includegraphics[scale=0.55]{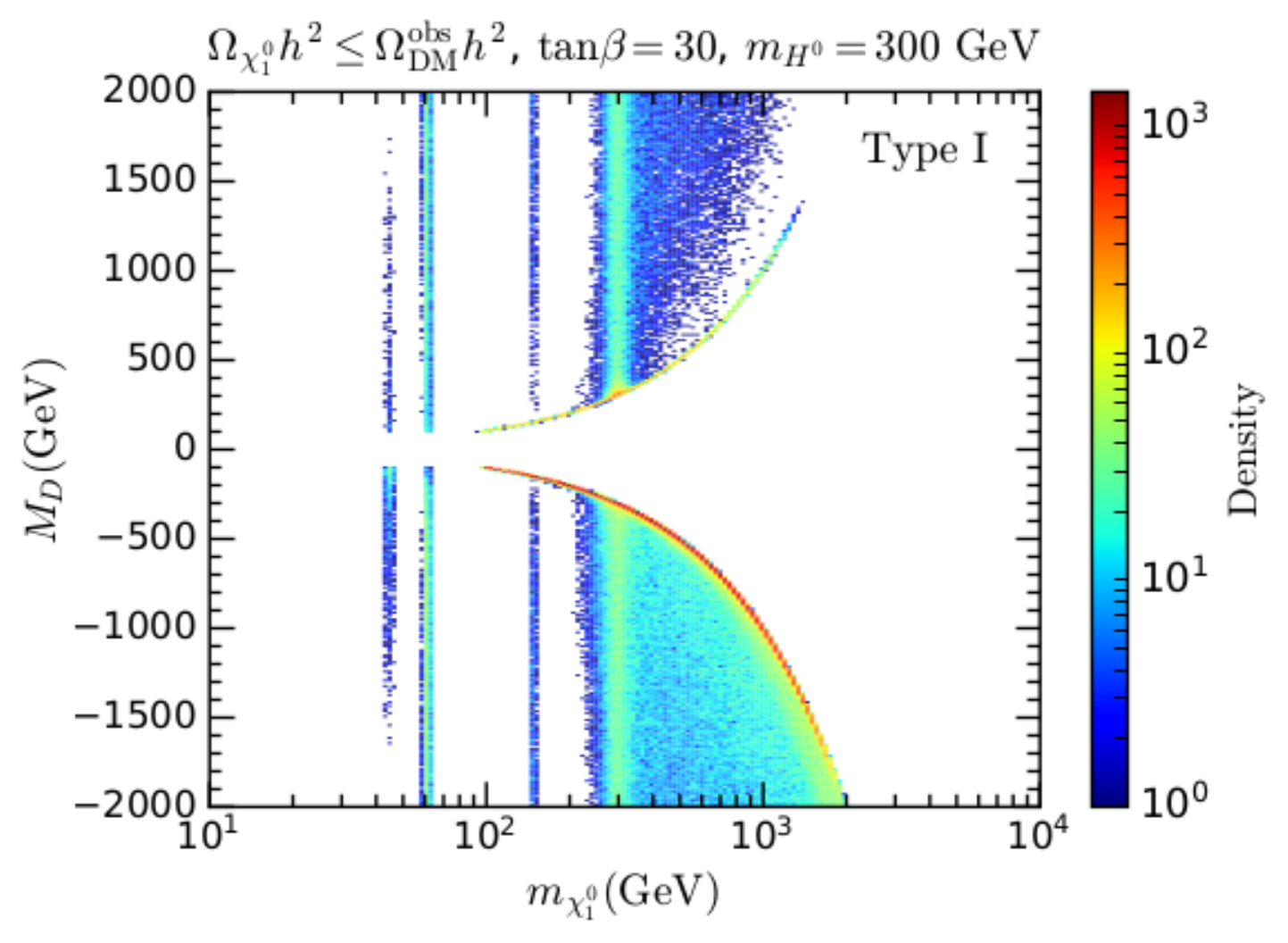}
    \caption{The same as Fig.~\ref{fig:type1H1}, but for the 
    Type I 2HDM with $\tanb=5$ (left panel) and $\tanb=30$ (right panel) with $\mH=300$ GeV. The rest of the parameters are specified in Table~\ref{tab:benchmarks}.}
      \label{fig:type1}
\end{figure}

\begin{figure}[ht]
    \centering
    \includegraphics[scale=0.55]{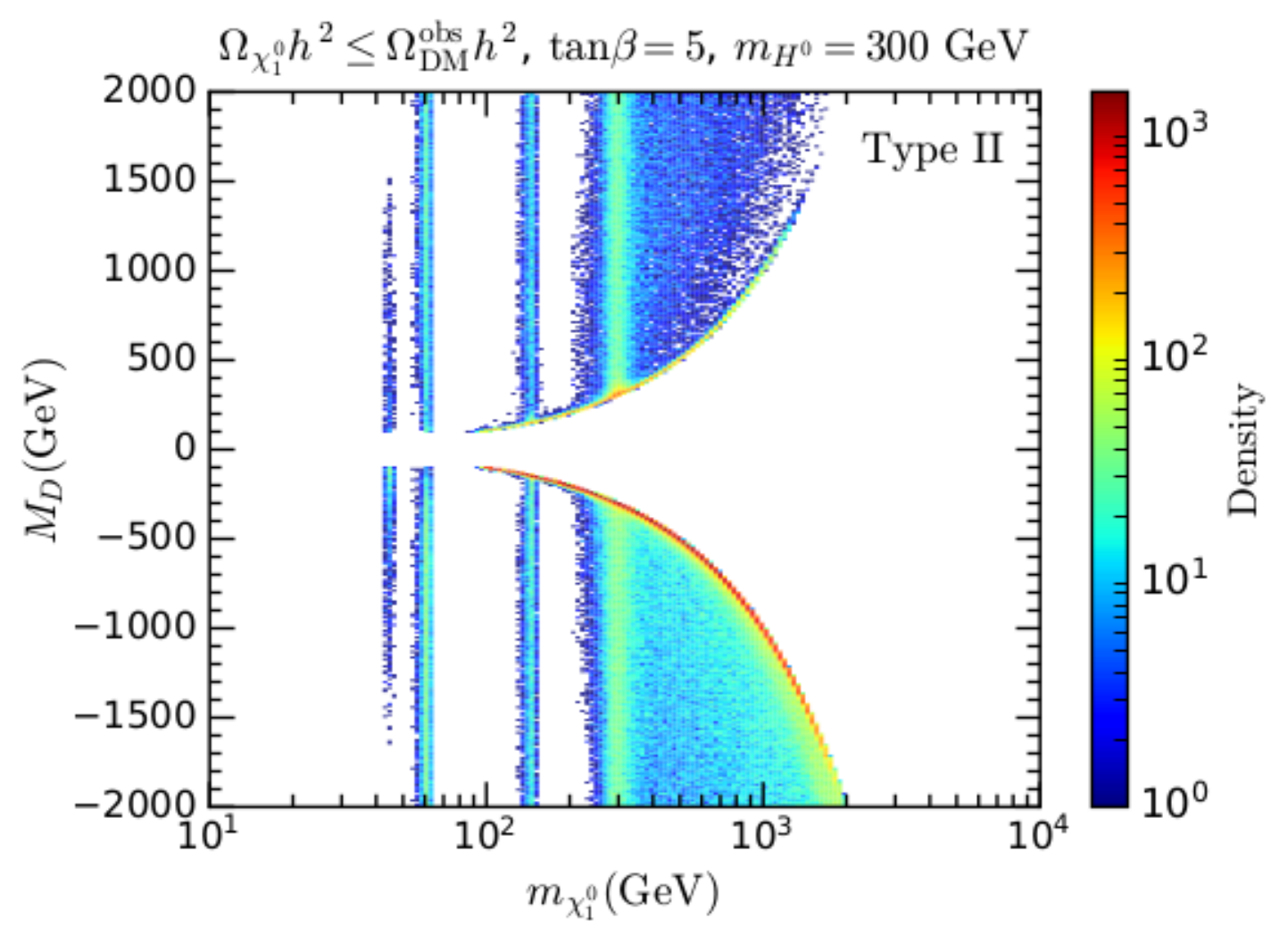}
    \includegraphics[scale=0.55]{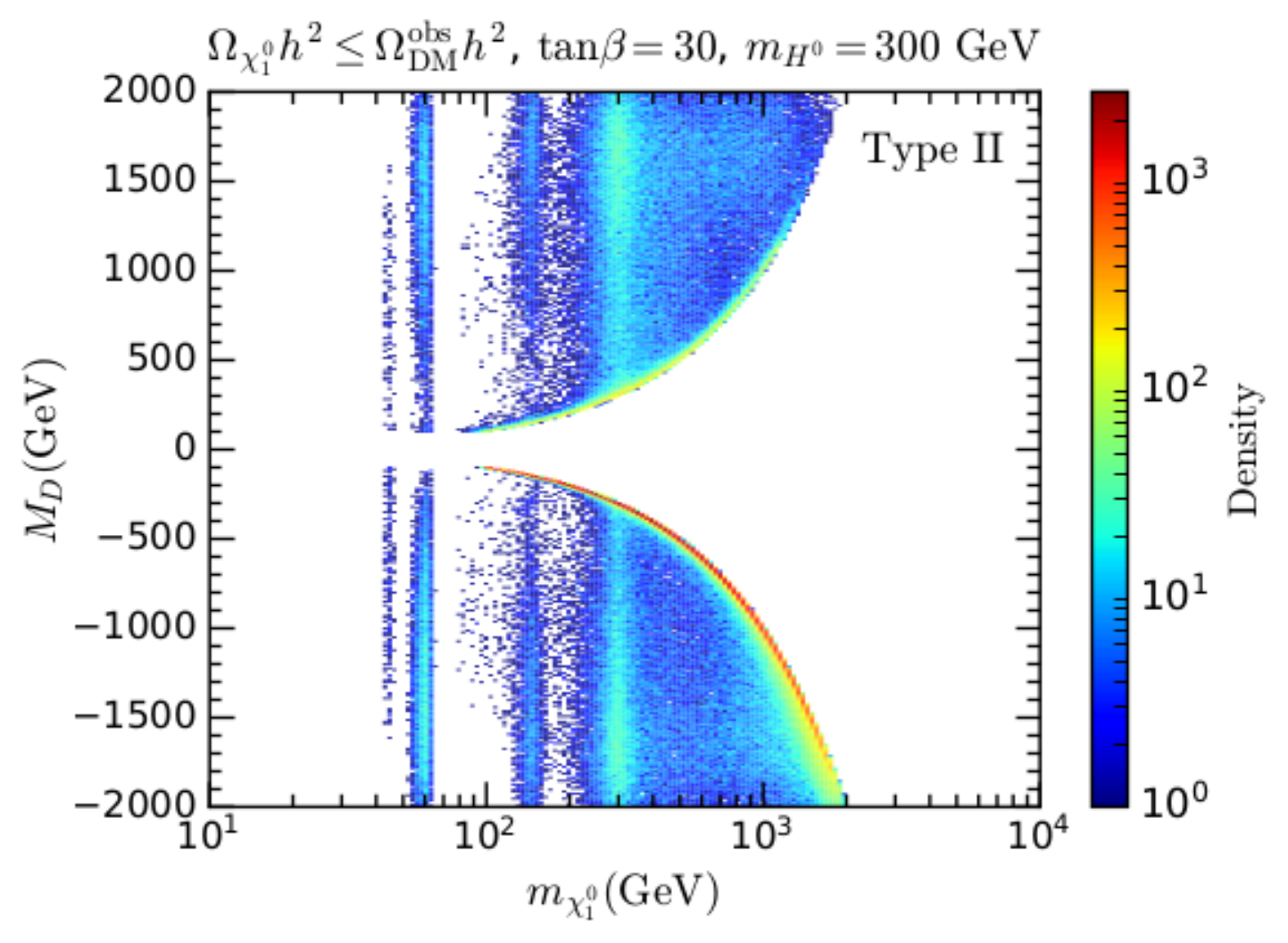}
   \caption{The same as Fig.~\ref{fig:type1} but for the Type II 2HDM.
   }
    \label{fig:type2}
\end{figure}

\begin{figure}[ht]
    \centering
    \includegraphics[scale=0.55]{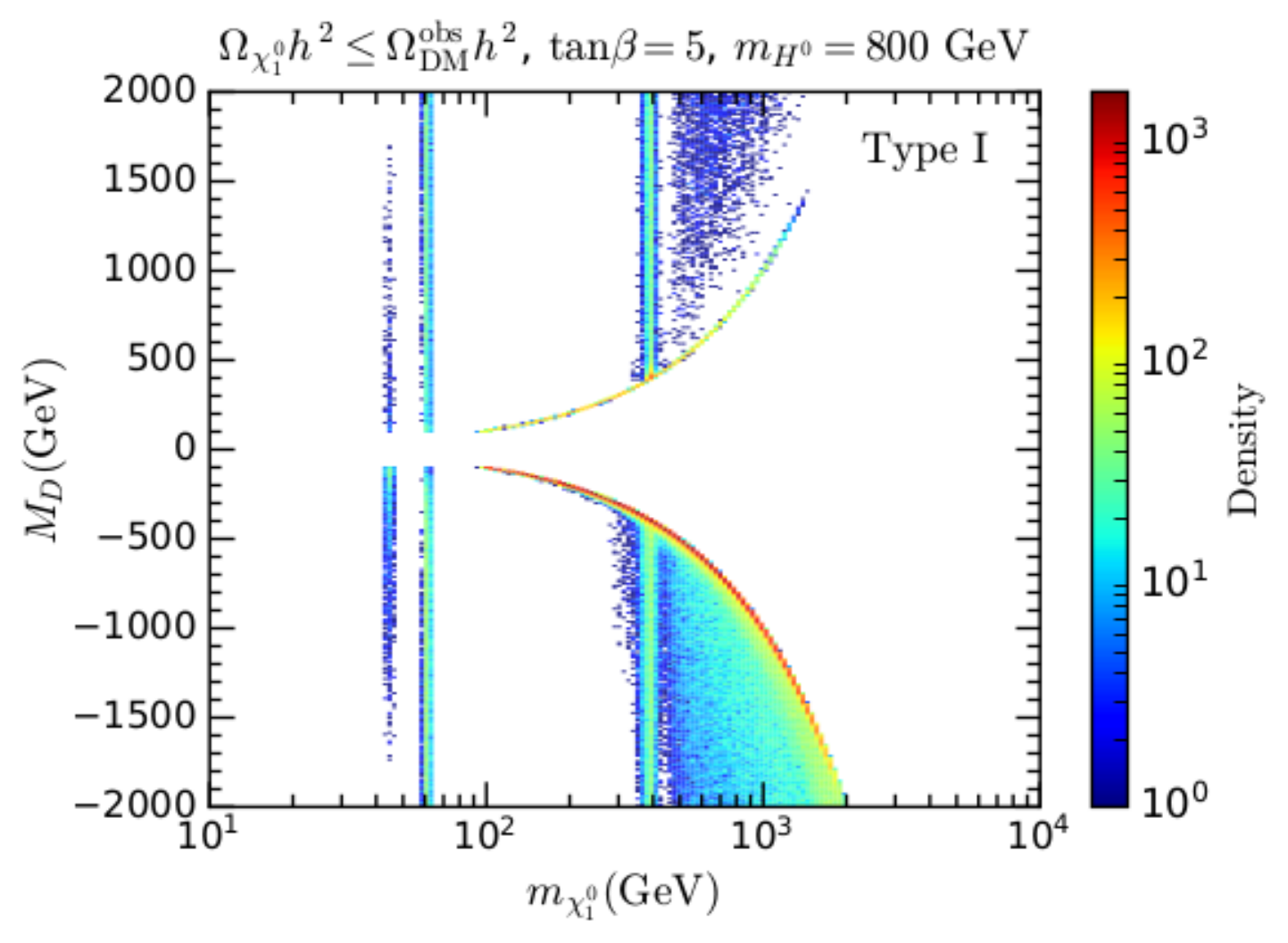}
    \includegraphics[scale=0.55]{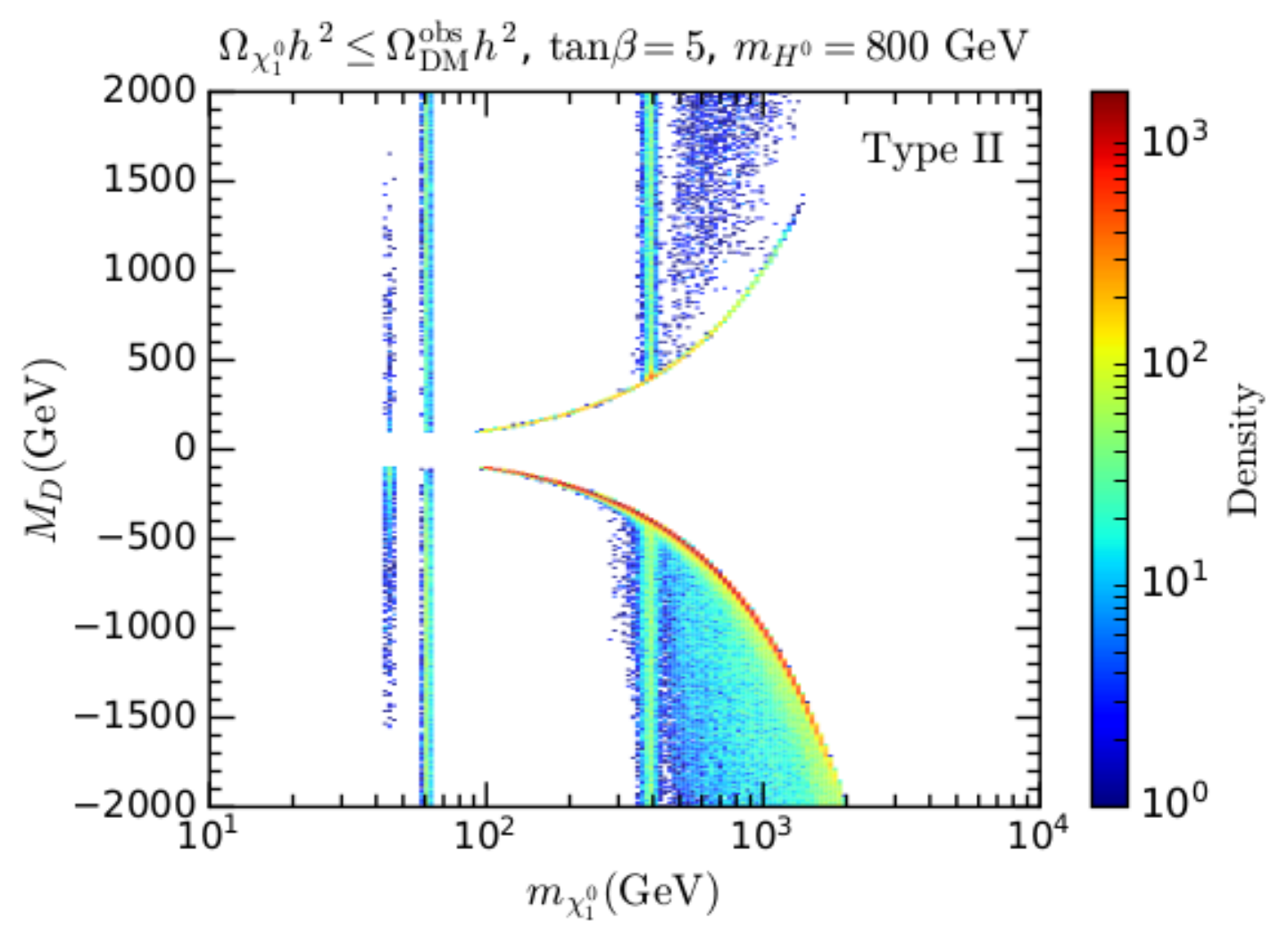}
    \caption{The same as Fig.~\ref{fig:type1H1}, but for the 
    Type I 2HDM with $\tanb=5$ (left panel) and  Type II $\tanb=5$ (right panel) with $\mH=800$ GeV. The rest of the parameters are  specified in Table~\ref{tab:benchmarks}.}
     \label{fig:H800}
\end{figure}

We illustrate these facts in Fig.~\ref{fig:type1}  (Type I),  Fig.~\ref{fig:type2} (Type II) and Fig.~\ref{fig:H800}, which are analogous to Fig.~\ref{fig:type1H1}, but for the benchmark models defined in Table~\ref{tab:benchmarks}. The only difference in the scan procedure is that now the $\tilde y_1, \tilde y_2$ couplings have been also surveyed, similarly to $y_1, y_2$, but in the [-1,1] range. 
The three figures show the dramatic enhancement of the regions of the parameter space consistent with the observed relic density and DD experiments, especially for $\mchi$ above the $\chi_1^0 \chi_1^0\rightarrow Z H^0$ threshold.

For all the benchmarks the values of $\mH$, $\mA$, $\mHc$ have been chosen to be in the safe region with respect to experimental constraints \cite{Bernon:2015qea}. It should be recalled here that their values, as well as that of $\tan\beta$, arise from the Higgs scalar potential, which we do not discuss here. For the purpose of this article they are free parameters.
Interestingly,  annihilation processes with $A$, $H^\pm$ in the final state can also be relevant for the relic density in some regions of the parameter space, even for the large masses considered here. 

The aforementioned points can be clearly appreciated for the Type I 2HDM in Fig.~\ref{fig:type2}. As in the case of a single Higgs in Fig.~\ref{fig:type1H1}, the region rescued for $\Omegachi = \Omegaobs$ (not plotted) is very similar to that of $\Omegachi \leq \Omegaobs$. Once more, the only difference between them are the narrow strips at $\mchi\simeq \pm M_D$, which are not especially dense regions for $\Omegachi = \Omegaobs$, except at the
``pure Higgsino" solution, $M_D\simeq \pm 1$ TeV. Aside from the various funnels visible in the plots at $\mchi\simeq m_Z, m_{h^0}, \mH,\dots$, 
all the allowed regions correspond to generalized blind spots, where $y^{\rm eff}_{\rm DD}$, as given by Eq.~(\ref{BS}), is nearly vanishing, though not necessarily by a cancellation between terms. As expected, the blind spot regions occur now for both positive and negative $M_D$, but interestingly there are still more solutions in the latter case. This is easily understood taking into account that in the Type I, the heavy Higgs contribution to $y^{\rm eff}_{\rm DD}$ is suppressed by the prefactor $\frac{m_h^2}{m_H^2} C_q =-\frac{m_h^2}{m_H^2}\cot\beta$. Then the light Higgs contribution, Eq.~(\ref{yhXXal}), must be small as well, which can be more easily achieved for $M_D<0$, as discussed in the previous subsection. Let us also mention that the allowed regions are very similar for both $\tan\beta = 5$ and $\tan\beta=30$.

The results for the Type II 2HDM, given in Fig.~\ref{fig:type2}, are very similar. The only noticeable difference is that for $\tan \beta=30$ the allowed region is larger than for the other cases and, furthermore, it is almost identical for positive and negative $M_D$. The reason is the following. In the Type II, the prefactor of the heavy Higgs contribution to $y^{\rm eff}_{\rm DD}$ reads $\frac{m_h^2}{m_H^2} C_q =\frac{m_h^2}{m_H^2}\tan\beta$ for the $d-$quarks. For $\tan \beta = 30$ this actually represents an enhancement, rather than a suppression. Hence, this term can be cancelled in Eq.~(\ref{BS}) with a sizeable light-Higgs contribution, and thus large $y-$couplings. In consequence, the processes $\chi_1^0\chi_1^0 \rightarrow h^0 \rightarrow {\rm SM\ SM}$, $\chi_1^0\chi_1^0 \rightarrow h^0\ h^0$ can be now efficient for DM annihilation. This especially happens for $\mchi\simgt 80$ GeV, i.e. above the $W^+W^-$ threshold. Likewise, since no small $y_{h \chi_i\chi_i}$ coupling is required now, the $M_D>0$, $M_D<0$ regions look alike.

The enhancement of the allowed regions holds even for rather large values of the extra Higgs states, especially above the mentioned $H^0Z$ threshold. This is illustrated in Fig.~\ref{fig:H800} for $\mH=m_A=m_{H^\pm}=800$ GeV and $\tan \beta =5$.

\begin{figure}[ht]
    \centering
    \includegraphics[scale=0.55]{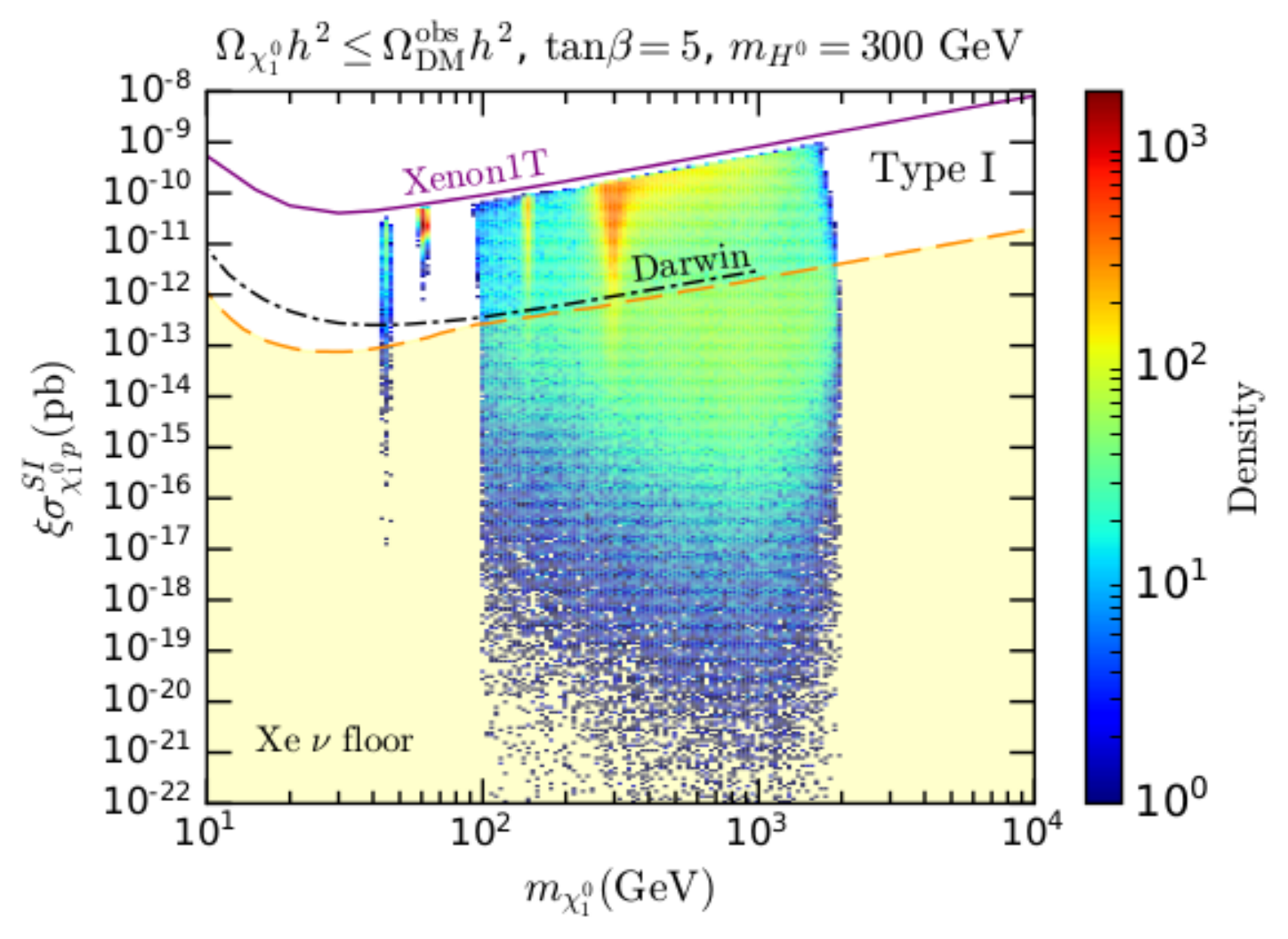}
    \caption{$\chi_1^0$-proton  spin-independent elastic  cross section, weighted by the scale factor $\xi$ for the Type I 2HDM benchmark with $\tan\beta=5$ and $\mH=300$ GeV. The color code indicates the relative density of points. Upper-bound lines from XENON1T \cite{Aprile:2018dbl} and sensitivity projections from the future DARWIN experiment \cite{Aalbers:2016jon} are also shown, as well as the neutrino floor for a xenon target \cite{Ruppin:2014bra}.}
     \label{fig:sigmaSI}
\end{figure}

 We have seen that the DD cross section can be suppressed in the 2HDM by a variety of mechanisms, not necessarily a cancellation between terms. However, it is still true that, in order to obtain extremely suppressed DD cross sections some kind of cancellation for $y^{\rm eff}_{\rm DD}$ is required. Consequently, the density of viable models is higher when the DD cross section is not much smaller than the future experimental constraints, as illustrated in Fig.~\ref{fig:sigmaSI}. This allows to be optimistic about the possibility that a scenario of the kind depicted in this paper might be detected by the next generation of direct detection experiments.

\vspace{0.2cm}

Finally, let us mention that there exist two additional 2HDMs, which are flavor changing neutral current (FCNC) free: the so-called X (or ``lepton-specific'') and Y (or ``flipped'') models. The corresponding $C_q$ factors are the same as those of the Type I and Type II, respectively, so the results presented in Figs. \ref{fig:type1}, \ref{fig:type2} and \ref{fig:H800}
apply to them as well.

\section{Summary and conclusions}
\label{sec:conclu}

$Z-$ and Higgs-portals are the most economical frameworks for WIMP dark matter. They are however under strong pressure (almost excluded), especially from direct detection experiments. One exception to this situation occurs when  dark matter particles  annihilate in a resonant way, i.e. the well-known $Z-$boson and Higgs funnels. Another way-out, equally interesting but not so explored, occurs when the spin-independent direct detection elastic cross section is suppressed due to some cancellation. These are the so-called ``blind spots" of the parameter space. Both, funnels and blind spots, require some degree of tuning and, indeed, only rather narrow regions of the parameter space can be rescued in this way.

In this paper we have focused on the structure of the blind spots when the Higgs sector is not minimal, as it happens in many BSM scenarios; more precisely, we have assumed a generic 2HDM. In addition, we have considered a dark sector consisting of a neutral fermion plus a Dirac doublet. The latter represents the minimal UV completion of a fermion-singlet Higgs-portal scenario for dark matter.
The funnel solutions change indeed little in this new framework, aside from the presence of additional funnels corresponding to the heavy Higgs and the pseudoscalar resonances. By contrast, the blind spot solutions change in a {\em qualitative} way, as discussed below.

In the first place, we have obtained general analytical expressions for the couplings of the dark matter to the light and heavy Higgses, which are the relevant ones for direct detection. This allowed us to write the effective coupling for direct detection, $y^{\rm eff}_{\rm DD}$, and thus the explicit condition for a blind spot.
In the case of a standard Higgs sector (which we re-visit as the decoupling limit of the 2HDM) the vanishing of $y^{\rm eff}_{\rm DD}$ implies the suppression of annihilation processes in the early Universe involving the Higgs, $\chi_1^0\chi_1^0 \rightarrow h^0 \rightarrow {\rm SM\ SM}$, $\chi_1^0\chi_1^0 \rightarrow h^0\ h^0$.
Then, dark matter annihilation can only occur thanks to a well-tempering mechanism, which implies an additional tuning. In particular, the masses of the singlet and the doublet must be rather close. In contrast, for a 2HDM, $y^{\rm eff}_{\rm DD}$ can be small because there is a  cancellation between the light and heavy Higgs  contributions. Hence,  the annihilation processes are not suppressed anymore, which enhances dramatically the allowed parameter space.

Actually, there is not even need of a cancellation between contributions. For example, for the Type I 2HDM the heavy Higgs contribution to $y^{\rm eff}_{\rm DD}$ is suppressed by its large mass, but also by an extra $\cot\beta$ factor. This means that the coupling of dark matter to the heavy Higgs can be large, making annihilation processes like  $\chi_1^0\chi_1^0 \rightarrow H^0\ H^0$, $\chi_1^0\chi_1^0 \rightarrow Z\ H^0$ efficient, and still keeping direct detection cross sections suppressed.

We have illustrated these facts in the alignment (without decoupling) limit for the possible FCNC-free 2HDMs, using representative benchmarks. Interestingly, the enhancement of the allowed parameter space is very important even for large, ${\cal O}(1\ {\rm TeV})$, masses of the extra Higgs states.

\vspace{0.3cm}
To summarize, the assumption of an extended Higgs sector has a great potential to rescue theoretically appealing WIMP scenarios.

\section*{Acknowledgments}
The work of JAC has been partially supported by Spanish Agencia Estatal de Investigaci\'on through the grant ``IFT Centro de Excelencia Severo Ochoa SEV-2016-0597’'and by MINECO project FPA 2016-78022-P.
The work of AD was partially supported by the National Science Foundation under grant PHY-1820860.
SR was supported by the Australian Research Council.
We thank Andrew Cheek for providing us with  the {\tt RAPIDD} tool tailored to  XENON1T.

\bibliographystyle{JHEP}  
\bibliography{references}

\providecommand{\href}[2]{#2}\begingroup\raggedright\begin{thebibliography}{10}

\bibitem{Bertone:2004pz}
G.~Bertone, D.~Hooper and J.~Silk, \emph{{Particle dark matter: Evidence,
  candidates and constraints}},
  \href{http://dx.doi.org/10.1016/j.physrep.2004.08.031}{\emph{Phys. Rept.}
  {\bf 405} (2005) 279--390}, [\href{http://arxiv.org/abs/hep-ph/0404175}{{\tt
  hep-ph/0404175}}].

\bibitem{Arcadi:2014lta}
G.~Arcadi, Y.~Mambrini and F.~Richard, \emph{{Z-portal dark matter}},
  \href{http://dx.doi.org/10.1088/1475-7516/2015/03/018}{\emph{JCAP} {\bf 1503}
  (2015) 018}, [\href{http://arxiv.org/abs/1411.2985}{{\tt 1411.2985}}].

\bibitem{Alves:2015pea}
A.~Alves, A.~Berlin, S.~Profumo and F.~S. Queiroz, \emph{{Dark Matter
  Complementarity and the Z$^\prime$ Portal}},
  \href{http://dx.doi.org/10.1103/PhysRevD.92.083004}{\emph{Phys. Rev.} {\bf
  D92} (2015) 083004}, [\href{http://arxiv.org/abs/1501.03490}{{\tt
  1501.03490}}].

\bibitem{Escudero:2016gzx}
M.~Escudero, A.~Berlin, D.~Hooper and M.-X. Lin, \emph{{Toward (Finally!)
  Ruling Out Z and Higgs Mediated Dark Matter Models}},
  \href{http://dx.doi.org/10.1088/1475-7516/2016/12/029}{\emph{JCAP} {\bf 1612}
  (2016) 029}, [\href{http://arxiv.org/abs/1609.09079}{{\tt 1609.09079}}].

\bibitem{Arcadi:2017kky}
G.~Arcadi, M.~Dutra, P.~Ghosh, M.~Lindner, Y.~Mambrini, M.~Pierre et~al.,
  \emph{{The waning of the WIMP? A review of models, searches, and
  constraints}},
  \href{http://dx.doi.org/10.1140/epjc/s10052-018-5662-y}{\emph{Eur. Phys. J.}
  {\bf C78} (2018) 203}, [\href{http://arxiv.org/abs/1703.07364}{{\tt
  1703.07364}}].

\bibitem{Gross:2017dan}
C.~Gross, O.~Lebedev and T.~Toma, \emph{{Cancellation Mechanism for
  Dark-Matter–Nucleon Interaction}},
  \href{http://dx.doi.org/10.1103/PhysRevLett.119.191801}{\emph{Phys. Rev.
  Lett.} {\bf 119} (2017) 191801}, [\href{http://arxiv.org/abs/1708.02253}{{\tt
  1708.02253}}].

\bibitem{Casas:2017jjg}
J.~A. Casas, D.~G. Cerdeño, J.~M. Moreno and J.~Quilis, \emph{{Reopening the
  Higgs portal for single scalar dark matter}},
  \href{http://dx.doi.org/10.1007/JHEP05(2017)036}{\emph{JHEP} {\bf 05} (2017)
  036}, [\href{http://arxiv.org/abs/1701.08134}{{\tt 1701.08134}}].

\bibitem{Baum:2017enm}
S.~Baum, M.~Carena, N.~R. Shah and C.~E.~M. Wagner, \emph{{Higgs portals for
  thermal Dark Matter. EFT perspectives and the NMSSM}},
  \href{http://dx.doi.org/10.1007/JHEP04(2018)069}{\emph{JHEP} {\bf 04} (2018)
  069}, [\href{http://arxiv.org/abs/1712.09873}{{\tt 1712.09873}}].

\bibitem{Arcadi:2019lka}
G.~Arcadi, A.~Djouadi and M.~Raidal, \emph{{Dark Matter through the Higgs
  portal}},  \href{http://arxiv.org/abs/1903.03616}{{\tt 1903.03616}}.

\bibitem{Cheung:2012qy}
C.~Cheung, L.~J. Hall, D.~Pinner and J.~T. Ruderman, \emph{{Prospects and Blind
  Spots for Neutralino Dark Matter}},
  \href{http://dx.doi.org/10.1007/JHEP05(2013)100}{\emph{JHEP} {\bf 05} (2013)
  100}, [\href{http://arxiv.org/abs/1211.4873}{{\tt 1211.4873}}].

\bibitem{Huang:2014xua}
P.~Huang and C.~E.~M. Wagner, \emph{{Blind Spots for neutralino Dark Matter in
  the MSSM with an intermediate $m_A$}},
  \href{http://dx.doi.org/10.1103/PhysRevD.90.015018}{\emph{Phys. Rev.} {\bf
  D90} (2014) 015018}, [\href{http://arxiv.org/abs/1404.0392}{{\tt
  1404.0392}}].

\bibitem{Anandakrishnan:2014fia}
A.~Anandakrishnan, B.~Shakya and K.~Sinha, \emph{{Dark matter at the
  pseudoscalar Higgs resonance in the phenomenological MSSM and SUSY GUTs}},
  \href{http://dx.doi.org/10.1103/PhysRevD.91.035029}{\emph{Phys. Rev.} {\bf
  D91} (2015) 035029}, [\href{http://arxiv.org/abs/1410.0356}{{\tt
  1410.0356}}].

\bibitem{Crivellin:2015bva}
A.~Crivellin, M.~Hoferichter, M.~Procura and L.~C. Tunstall, \emph{{Light
  stops, blind spots, and isospin violation in the MSSM}},
  \href{http://dx.doi.org/10.1007/JHEP07(2015)129}{\emph{JHEP} {\bf 07} (2015)
  129}, [\href{http://arxiv.org/abs/1503.03478}{{\tt 1503.03478}}].

\bibitem{Cheung:2013dua}
C.~Cheung and D.~Sanford, \emph{{Simplified Models of Mixed Dark Matter}},
  \href{http://dx.doi.org/10.1088/1475-7516/2014/02/011}{\emph{JCAP} {\bf 1402}
  (2014) 011}, [\href{http://arxiv.org/abs/1311.5896}{{\tt 1311.5896}}].

\bibitem{Berlin:2015wwa}
A.~Berlin, S.~Gori, T.~Lin and L.-T. Wang, \emph{{Pseudoscalar Portal Dark
  Matter}}, \href{http://dx.doi.org/10.1103/PhysRevD.92.015005}{\emph{Phys.
  Rev.} {\bf D92} (2015) 015005}, [\href{http://arxiv.org/abs/1502.06000}{{\tt
  1502.06000}}].

\bibitem{Banerjee:2016hsk}
S.~Banerjee, S.~Matsumoto, K.~Mukaida and Y.-L.~S. Tsai, \emph{{WIMP Dark
  Matter in a Well-Tempered Regime: A case study on Singlet-Doublets Fermionic
  WIMP}}, \href{http://dx.doi.org/10.1007/JHEP11(2016)070}{\emph{JHEP} {\bf 11}
  (2016) 070}, [\href{http://arxiv.org/abs/1603.07387}{{\tt 1603.07387}}].

\bibitem{Arcadi:2018pfo}
G.~Arcadi, \emph{{2HDM portal for Singlet-Doublet Dark Matter}},
  \href{http://dx.doi.org/10.1140/epjc/s10052-018-6327-6}{\emph{Eur. Phys. J.}
  {\bf C78} (2018) 864}, [\href{http://arxiv.org/abs/1804.04930}{{\tt
  1804.04930}}].

\bibitem{Branco:2011iw}
G.~C. Branco, P.~M. Ferreira, L.~Lavoura, M.~N. Rebelo, M.~Sher and J.~P.
  Silva, \emph{{Theory and phenomenology of two-Higgs-doublet models}},
  \href{http://dx.doi.org/10.1016/j.physrep.2012.02.002}{\emph{Phys. Rept.}
  {\bf 516} (2012) 1--102}, [\href{http://arxiv.org/abs/1106.0034}{{\tt
  1106.0034}}].

\bibitem{Jiang:2019soj}
X.-M. Jiang, C.~Cai, Z.-H. Yu, Y.-P. Zeng and H.-H. Zhang,
  \emph{{Pseudo-Nambu-Goldstone dark matter and two-Higgs-doublet models}},
  \href{http://dx.doi.org/10.1103/PhysRevD.100.075011}{\emph{Phys. Rev.} {\bf
  D100} (2019) 075011}, [\href{http://arxiv.org/abs/1907.09684}{{\tt
  1907.09684}}].

\bibitem{Bernon:2015qea}
J.~Bernon, J.~F. Gunion, H.~E. Haber, Y.~Jiang and S.~Kraml,
  \emph{{Scrutinizing the alignment limit in two-Higgs-doublet models:
  m$_h$=125 GeV}},
  \href{http://dx.doi.org/10.1103/PhysRevD.92.075004}{\emph{Phys. Rev.} {\bf
  D92} (2015) 075004}, [\href{http://arxiv.org/abs/1507.00933}{{\tt
  1507.00933}}].

\bibitem{Aad:2012tfa}
{\scshape ATLAS} collaboration, G.~Aad et~al., \emph{{Observation of a new
  particle in the search for the Standard Model Higgs boson with the ATLAS
  detector at the LHC}},
  \href{http://dx.doi.org/10.1016/j.physletb.2012.08.020}{\emph{Phys.Lett.}
  {\bf B716} (2012) 1--29}, [\href{http://arxiv.org/abs/1207.7214}{{\tt
  1207.7214}}].

\bibitem{Chatrchyan:2012ufa}
{\scshape CMS} collaboration, S.~Chatrchyan et~al., \emph{{Observation of a new
  boson at a mass of 125 GeV with the CMS experiment at the LHC}},
  \href{http://dx.doi.org/10.1016/j.physletb.2012.08.021}{\emph{Phys.Lett.}
  {\bf B716} (2012) 30--61}, [\href{http://arxiv.org/abs/1207.7235}{{\tt
  1207.7235}}].

\bibitem{Aad:2013wqa}
{\scshape ATLAS} collaboration, G.~Aad et~al., \emph{{Measurements of Higgs
  boson production and couplings in diboson final states with the ATLAS
  detector at the LHC}},
  \href{http://dx.doi.org/10.1016/j.physletb.2014.05.011,
  10.1016/j.physletb.2013.08.010}{\emph{Phys. Lett.} {\bf B726} (2013)
  88--119}, [\href{http://arxiv.org/abs/1307.1427}{{\tt 1307.1427}}].

\bibitem{Chatrchyan:2013lba}
{\scshape CMS} collaboration, S.~Chatrchyan et~al., \emph{{Observation of a new
  boson with mass near 125 GeV in pp collisions at $\sqrt{s}$ = 7 and 8 TeV}},
  \href{http://dx.doi.org/10.1007/JHEP06(2013)081}{\emph{JHEP} {\bf 06} (2013)
  081}, [\href{http://arxiv.org/abs/1303.4571}{{\tt 1303.4571}}].

\bibitem{Sirunyan:2017exp}
{\scshape CMS} collaboration, A.~M. Sirunyan et~al., \emph{{Measurements of
  properties of the Higgs boson decaying into the four-lepton final state in pp
  collisions at $ \sqrt{s}=13 $ TeV}},
  \href{http://dx.doi.org/10.1007/JHEP11(2017)047}{\emph{JHEP} {\bf 11} (2017)
  047}, [\href{http://arxiv.org/abs/1706.09936}{{\tt 1706.09936}}].

\bibitem{Aaboud:2018wps}
{\scshape ATLAS} collaboration, M.~Aaboud et~al., \emph{{Measurement of the
  Higgs boson mass in the $H\rightarrow ZZ^* \rightarrow 4\ell$ and $H
  \rightarrow \gamma\gamma$ channels with $\sqrt{s}=13$ TeV $pp$ collisions
  using the ATLAS detector}},
  \href{http://dx.doi.org/10.1016/j.physletb.2018.07.050}{\emph{Phys. Lett.}
  {\bf B784} (2018) 345--366}, [\href{http://arxiv.org/abs/1806.00242}{{\tt
  1806.00242}}].

\bibitem{Aprile:2018dbl}
{\scshape XENON} collaboration, E.~Aprile et~al., \emph{{Dark Matter Search
  Results from a One Ton-Year Exposure of XENON1T}},
  \href{http://dx.doi.org/10.1103/PhysRevLett.121.111302}{\emph{Phys. Rev.
  Lett.} {\bf 121} (2018) 111302}, [\href{http://arxiv.org/abs/1805.12562}{{\tt
  1805.12562}}].

\bibitem{Aprile:2019dbj}
{\scshape XENON} collaboration, E.~Aprile et~al., \emph{{Constraining the
  spin-dependent WIMP-nucleon cross sections with XENON1T}},
  \href{http://dx.doi.org/10.1103/PhysRevLett.122.141301}{\emph{Phys. Rev.
  Lett.} {\bf 122} (2019) 141301}, [\href{http://arxiv.org/abs/1902.03234}{{\tt
  1902.03234}}].

\bibitem{Amole:2019fdf}
{\scshape PICO} collaboration, C.~Amole et~al., \emph{{Dark Matter Search
  Results from the Complete Exposure of the PICO-60 C$_3$F$_8$ Bubble
  Chamber}}, \href{http://dx.doi.org/10.1103/PhysRevD.100.022001}{\emph{Phys.
  Rev.} {\bf D100} (2019) 022001}, [\href{http://arxiv.org/abs/1902.04031}{{\tt
  1902.04031}}].

\bibitem{ArkaniHamed:2006mb}
N.~Arkani-Hamed, A.~Delgado and G.~F. Giudice, \emph{{The Well-tempered
  neutralino}},
  \href{http://dx.doi.org/10.1016/j.nuclphysb.2006.02.010}{\emph{Nucl. Phys.}
  {\bf B741} (2006) 108--130}, [\href{http://arxiv.org/abs/hep-ph/0601041}{{\tt
  hep-ph/0601041}}].

\bibitem{Heister:2003zk}
{\scshape ALEPH} collaboration, A.~Heister et~al., \emph{{Absolute mass lower
  limit for the lightest neutralino of the MSSM from e+ e- data at s**(1/2) up
  to 209-GeV}},
  \href{http://dx.doi.org/10.1016/j.physletb.2003.12.066}{\emph{Phys. Lett.}
  {\bf B583} (2004) 247--263}.

\bibitem{Christensen:2008py}
N.~D. Christensen and C.~Duhr, \emph{{FeynRules - Feynman rules made easy}},
  \href{http://dx.doi.org/10.1016/j.cpc.2009.02.018}{\emph{Comput. Phys.
  Commun.} {\bf 180} (2009) 1614--1641},
  [\href{http://arxiv.org/abs/0806.4194}{{\tt 0806.4194}}].

\bibitem{Alloul:2013bka}
A.~Alloul, N.~D. Christensen, C.~Degrande, C.~Duhr and B.~Fuks,
  \emph{{FeynRules 2.0 - A complete toolbox for tree-level phenomenology}},
  \href{http://dx.doi.org/10.1016/j.cpc.2014.04.012}{\emph{Comput. Phys.
  Commun.} {\bf 185} (2014) 2250--2300},
  [\href{http://arxiv.org/abs/1310.1921}{{\tt 1310.1921}}].

\bibitem{Christensen:2009jx}
N.~D. Christensen, P.~de~Aquino, C.~Degrande, C.~Duhr, B.~Fuks, M.~Herquet
  et~al., \emph{{A Comprehensive approach to new physics simulations}},
  \href{http://dx.doi.org/10.1140/epjc/s10052-011-1541-5}{\emph{Eur. Phys. J.}
  {\bf C71} (2011) 1541}, [\href{http://arxiv.org/abs/0906.2474}{{\tt
  0906.2474}}].

\bibitem{Degrande:2014vpa}
C.~Degrande, \emph{{Automatic evaluation of UV and R2 terms for beyond the
  Standard Model Lagrangians: a proof-of-principle}},
  \href{http://dx.doi.org/10.1016/j.cpc.2015.08.015}{\emph{Comput. Phys.
  Commun.} {\bf 197} (2015) 239--262},
  [\href{http://arxiv.org/abs/1406.3030}{{\tt 1406.3030}}].

\bibitem{Belanger:2006is}
G.~Belanger, F.~Boudjema, A.~Pukhov and A.~Semenov, \emph{{MicrOMEGAs 2.0: A
  Program to calculate the relic density of dark matter in a generic model}},
  \href{http://dx.doi.org/10.1016/j.cpc.2006.11.008}{\emph{Comput. Phys.
  Commun.} {\bf 176} (2007) 367--382},
  [\href{http://arxiv.org/abs/hep-ph/0607059}{{\tt hep-ph/0607059}}].

\bibitem{Feroz:2007kg}
F.~Feroz and M.~P. Hobson, \emph{{Multimodal nested sampling: an efficient and
  robust alternative to MCMC methods for astronomical data analysis}},
  \href{http://dx.doi.org/10.1111/j.1365-2966.2007.12353.x}{\emph{Mon. Not.
  Roy. Astron. Soc.} {\bf 384} (2008) 449},
  [\href{http://arxiv.org/abs/0704.3704}{{\tt 0704.3704}}].

\bibitem{Feroz:2008xx}
F.~Feroz, M.~P. Hobson and M.~Bridges, \emph{{MultiNest: an efficient and
  robust Bayesian inference tool for cosmology and particle physics}},
  \href{http://dx.doi.org/10.1111/j.1365-2966.2009.14548.x}{\emph{Mon. Not.
  Roy. Astron. Soc.} {\bf 398} (2009) 1601--1614},
  [\href{http://arxiv.org/abs/0809.3437}{{\tt 0809.3437}}].

\bibitem{Feroz:2013hea}
F.~Feroz, M.~P. Hobson, E.~Cameron and A.~N. Pettitt, \emph{{Importance Nested
  Sampling and the MultiNest Algorithm}},
  \href{http://arxiv.org/abs/1306.2144}{{\tt 1306.2144}}.

\bibitem{deAustri:2006jwj}
R.~Ruiz~de Austri, R.~Trotta and L.~Roszkowski, \emph{{A Markov chain Monte
  Carlo analysis of the CMSSM}},
  \href{http://dx.doi.org/10.1088/1126-6708/2006/05/002}{\emph{JHEP} {\bf 05}
  (2006) 002}, [\href{http://arxiv.org/abs/hep-ph/0602028}{{\tt
  hep-ph/0602028}}].

\bibitem{Aghanim:2018eyx}
{\scshape Planck} collaboration, N.~Aghanim et~al., \emph{{Planck 2018 results.
  VI. Cosmological parameters}},  \href{http://arxiv.org/abs/1807.06209}{{\tt
  1807.06209}}.

\bibitem{Cerdeno:2018bty}
D.~G. Cerdeño, A.~Cheek, E.~Reid and H.~Schulz, \emph{{Surrogate Models for
  Direct Dark Matter Detection}},
  \href{http://dx.doi.org/10.1088/1475-7516/2018/08/011}{\emph{JCAP} {\bf 1808}
  (2018) 011}, [\href{http://arxiv.org/abs/1802.03174}{{\tt 1802.03174}}].

\bibitem{Carena:2013ooa}
M.~Carena, I.~Low, N.~R. Shah and C.~E.~M. Wagner, \emph{{Impersonating the
  Standard Model Higgs Boson: Alignment without Decoupling}},
  \href{http://dx.doi.org/10.1007/JHEP04(2014)015}{\emph{JHEP} {\bf 04} (2014)
  015}, [\href{http://arxiv.org/abs/1310.2248}{{\tt 1310.2248}}].

\bibitem{Aalbers:2016jon}
{\scshape DARWIN} collaboration, J.~Aalbers et~al., \emph{{DARWIN: towards the
  ultimate dark matter detector}},
  \href{http://dx.doi.org/10.1088/1475-7516/2016/11/017}{\emph{JCAP} {\bf 1611}
  (2016) 017}, [\href{http://arxiv.org/abs/1606.07001}{{\tt 1606.07001}}].

\bibitem{Ruppin:2014bra}
F.~Ruppin, J.~Billard, E.~Figueroa-Feliciano and L.~Strigari,
  \emph{{Complementarity of dark matter detectors in light of the neutrino
  background}}, \href{http://dx.doi.org/10.1103/PhysRevD.90.083510}{\emph{Phys.
  Rev.} {\bf D90} (2014) 083510}, [\href{http://arxiv.org/abs/1408.3581}{{\tt
  1408.3581}}].

\end{thebibliography}\endgroup

\end{document}